\documentclass{aa}
\usepackage{graphicx}
\usepackage{txfonts}
\usepackage{natbib}
\bibpunct{(}{)}{;}{a}{}{,} 
\newcommand{\mincir}{\raise -2.truept\hbox{\rlap{\hbox{$\sim$}}\raise5.truept
\hbox{$<$}\ }}
\newcommand{\magcir}{\raise -2.truept\hbox{\rlap{\hbox{$\sim$}}\raise5.truept
\hbox{$>$}\ }}
\newcommand{\siml}{\raise -2.truept\hbox{\rlap{\hbox{$\sim$}}\raise5.truept
\hbox{$<$}\ }}
\newcommand{\simg}{\raise -2.truept\hbox{\rlap{\hbox{$\sim$}}\raise5.truept
\hbox{$>$}\ }}
\newcommand{\be}{\begin{equation}}
\newcommand{\ee}{\end{equation}}
\newcommand{\ba}{\begin{eqnarray}}
\newcommand{\ea}{\end{eqnarray}}
\newcommand{\ks}{\mathrm{km\,s^{-1}}}
\newcommand{\ksmpc}{\mathrm{km\,s^{-1}\,Mpc^{-1}}}
\newcommand{\kpc}{\mathrm{kpc}}
\newcommand{\mpc}{\mathrm{Mpc}}
\newcommand{\ma}{\mathrm{mag}}
\newcommand{\maa}{\mathrm{mag\,arcsec^{-2}}}
\newcommand{\gmaa}{g\,\mathrm{mag\,arcsec^{-2}}}

\newcommand{\arcminn}{\mathrm{arcmin^2}}
\newcommand{\kev}{\mathrm{keV}}
\newcommand{\se}{\mathrm{s}}

\newcommand{\lsun}{L_{\odot}}
\newcommand{\mtre}{\times 10^{13}~M_{\odot}}
\newcommand{\mqua}{\times 10^{14}~M_{\odot}}

\newcommand{\ml}{M_{\odot}/L_{\odot}}
\newcommand{\degree}{\ensuremath{\mathrm{^\circ}}}
\newcommand{\arcmm}{\ensuremath{\mathrm{^\prime}\;}}
\newcommand{\arcm}{\ensuremath{\mathrm{^\prime}}}

\newcommand{\arcs}{\ensuremath{\arcm\hskip -0.1em\arcm}}
\newcommand{\dotarcs}{\,\rlap{\hbox{$\mathrm{^\prime\hskip-0.1em^\prime}$}}{\hbox{$.$}}\,}

\newcommand{\dotsec}{\,\rlap{\hbox{$\mathrm{^s}$}}{\hbox{$.$}}\,}

\begin{document}

\title{A newly identified galaxy group thanks to tidal streams of intragroup  light \thanks{Full Table 1 is only available in electronic form
at the CDS via anonymous ftp to  cdsarc.cds.unistra.fr (130.79.128.5)
or via https://cdsarc.cds.unistra.fr/cgi-bin/qcat?J/A+A/}}

%

\author{M. Girardi\inst{1,2},
S. Zarattini\inst{3,4},
W. Boschin\inst{5,2,6,7},
M. Nonino\inst{2},
I. Bartalucci \inst{8},
A. Mercurio\inst{9,10},
N. Nocerino\inst{1},
P. Rosati\inst{11}}


  \institute{Dipartimento di Fisica dell'Universit\`a degli Studi di Trieste -
Sezione di Astronomia, via Tiepolo 11, I-34143 Trieste, Italy \email{marisa.girardi@inaf.it}
\and INAF - Osservatorio Astronomico di Trieste, via Tiepolo 11,
I-34143 Trieste, Italy
\and Dipartimento di Fisica e Astronomia ``G. Galilei'', Universit\`a di Padova, vicolo dell'Osservatorio 3, I-35122 Padova, Italy
\and INAF - Osservatorio Astronomico di Padova, vicolo dell'Osservatorio 2, I-35122 Padova, Italy
\and Fundaci\'on Galileo Galilei - INAF (Telescopio Nazionale
  Galileo), Rambla Jos\'e Ana Fern\'andez Perez 7, E-38712 Bre\~na
  Baja (La Palma), Canary Islands, Spain
\and Instituto de Astrof\'{\i}sica de Canarias, C/V\'{\i}a L\'actea
s/n, E-38205 La Laguna (Tenerife), Canary Islands, Spain
\and Departamento de Astrof\'{\i}sica, Univ. de La Laguna, Av. del
Astrof\'{\i}sico Francisco S\'anchez s/n, E-38205 La Laguna
(Tenerife), Spain
\and INAF - Istituto di Astrofisica Spaziale e Fisica Cosmica di Milano, Via A. Corti 12, I-20133 Milano, Italy
\and Dipartimento di Fisica E.R. Caianiello, Universit\`a Degli Studi di Salerno, Via Giovanni Paolo II, I-84084 Fisciano (SA), Italy
\and
INAF - Osservatorio Astronomico di Capodimonte, via Moiariello 16, I-80131 Napoli, Italy
\and Dipartimento di Fisica e Scienze della Terra, Universit\`a  di Ferrara, via Saragat 1, I-44122 Ferrara, Italy
  }

  \date{Received  / Accepted }

  \abstract {In the accretion-driven growth scenario, part of the
    intracluster light is formed in the group environment.}  {We
    report the serendipitous discovery of a group of galaxies with
    signs of diffuse light in the foreground of the known galaxy
    cluster MACS\,J0329-0211 at $z \sim 0.45$.} {Our investigation
    began with the detection of diffuse light streams around a pair of
    bright galaxies in the southeastern region of a Suprime-Cam image
    of the galaxy cluster MACS\,J0329-0211. Our analysis is based on
    the extended CLASH-VLT redshift catalog and on new spectroscopic
    data obtained ad hoc with the Italian Telescopio Nazionale {\em
      Galileo}. We use the density reconstruction method to analyze
    the redshift distribution of the galaxies in the region around the
    galaxy pair. We also use available photometric and X-ray data to
    better characterize the properties of the group.}  {Thanks to the
    large amount of redshift data collected in this region, we have
    been able to discover the existence of a group of galaxies, here
    called GrG\,J0330-0218, which is associated with the pair of
    galaxies. These are the two brightest group galaxies (BGG1 and
    BGG2).  We extracted 41 group members from the redshift catalog
    and estimate a mean redshift $z = 0.1537$ and a line-of-sight
    velocity dispersion $\sigma_{\rm V} \sim 370~\ks$.  In the
    phase-space diagram, the distribution of the galaxies of
    GrG\,J0330-0218 follows the characteristic trumpet-shaped pattern,
    which is related to the escape velocity of galaxy clusters,
    suggesting that the group is a virialized structure. Under this
    assumption, the mass of the group is $M_{200}\sim 6~\mtre$. We
    also measured a mass-to-light ratio of $\sim 130~\ml$ and a
    luminosity fraction of diffuse light of $\sim 20\%$ within
    $0.5~R_{200}$.}  {We conjecture that galaxy pairs that are
    surrounded by diffuse light, probably due to tidal interactions,
    can serve as signposts for groups.}


\titlerunning{A newly identified group and intragroup light} 
\authorrunning{Girardi et al.}

\maketitle

\section{Introduction}
\label{intro}

\begin{figure*}[!ht]
\centering
\resizebox{\hsize}{!}{\includegraphics[width=18cm]{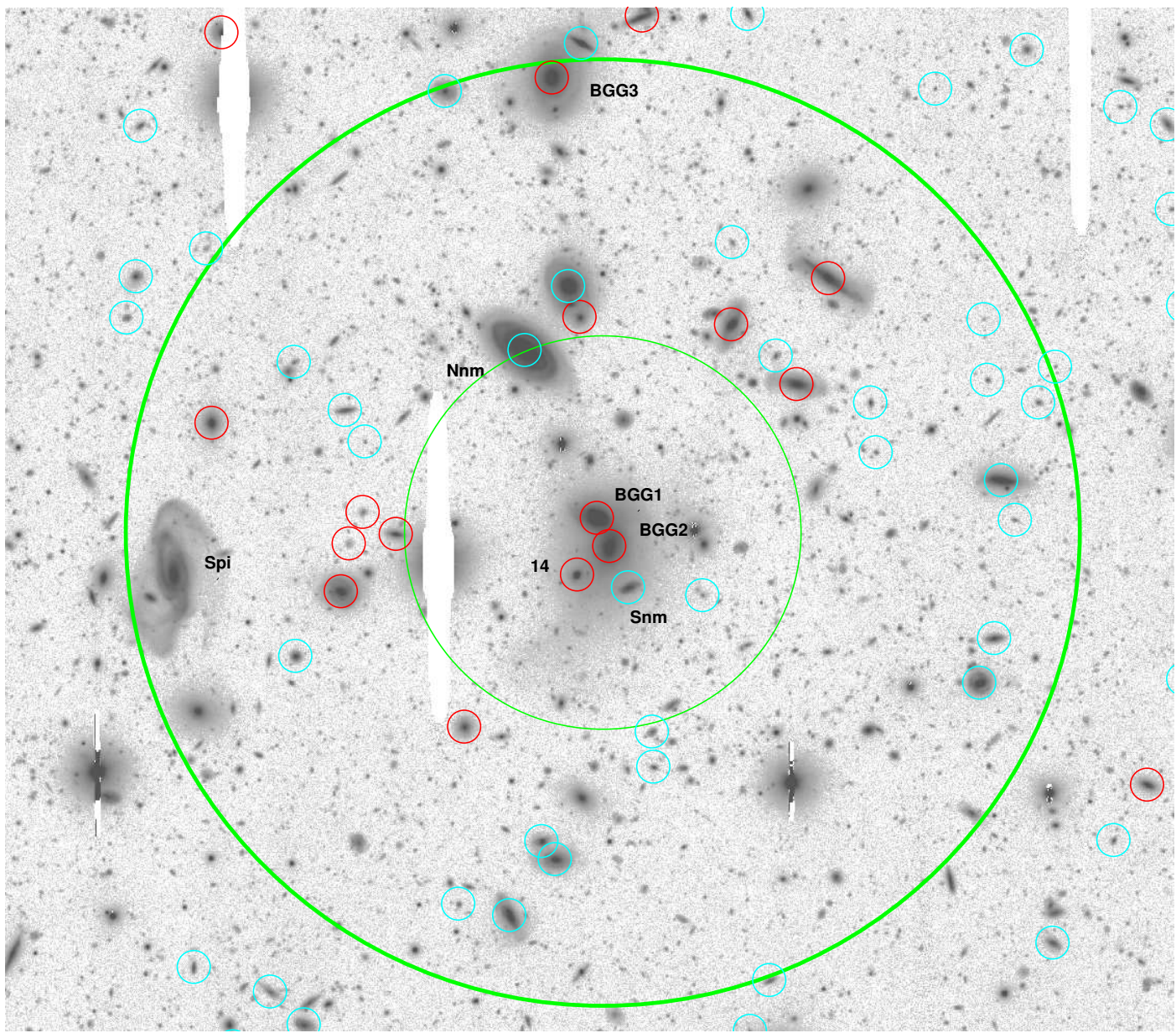}}
\caption{ Southeastern region of the Suprime-Cam $R_{\rm C}$-band
  image of the cluster MACS0329 (north top and east left) showing the
  group region within $0.5~R_{200} \sim 2.4\arcmin$ (enclosed in the
  large, thick green circle).  The thin green circle indicates the
  $1\arcmin$ radius region ($1\arcmin \sim 160~\kpc$  at the group redshift).
  Galaxies with available redshift are highlighted with small circles,
  in particular red or cyan circles in the case of members or
  nonmembers according to the selection procedure (see
  Sects.~\ref{data} and \ref{memb}). Labels indicate objects discussed
  in the text.}
\label{figottico}
\end{figure*}

Galaxy clusters evolve and increase in mass through a hierarchical
merging process from poor groups to rich clusters. As shown by
numerical simulations, accretion of smaller systems at the galaxy or
group scale is the main channel of cluster growth (e.g.,
\citealt{benavides2020} and references therein). In this scenario,
some of the intracluster diffuse light could originate from the group
environment.

Diffuse light is the light from stars that float freely in the
gravitational potential of galaxy systems and that are not bound to a
galaxy. For simplicity, we use the term ICL when referring to both
intragroup and intracluster light.  Diffuse light in galaxy systems
was first discovered by \citet{zwicky1951} in the Coma cluster and
studied in the following years. We refer to \citet{contini2021} and
\citet{montes2022} as two very recent reviews on this topic.  
  Most of the ICL in clusters is concentrated around the brightest
cluster galaxy (BCG) where it is considered to be an extended
component over the S\'ersic light profile of the BCG itself.

To detect ICL there are three methods used in the literature.
  The most commonly used method for calculating the ICL is that based
  on a surface brightness cut (e.g., \citealt{feldmeir2004};
  \citealt{burke2012}; \citealt{furnell2021}). It assumes that all
  light below a certain surface brightness can be fully attributed to
  the ICL.  The disadvantage of this method is that images with
  different exposure times for the same galaxy cluster give different
  amounts of the ICL (e.g., \citealt{montes2018}). Moreover, it does
  not account for the ICL that overlaps with the BCG in the transition
  region. As an alternative method, BCG and ICL light can be described
  by adding separate models for the two components.  This method based
  on the surface brightness profile fit can be heavily degenerate
  (e.g., \citealt{janowiecki2010}). Moreover, as pointed out by
  \citet{montes2018}, it also fails to account for ICL not
  concentrated around the BCG, asymmetries in the light distribution,
  and substructures such as tidal streams. The third approach is based
  on 2D techniques, such as fitting algorithms to take into account
  most of the galaxies in the cluster (e.g., \citealt{giallongo2014};
  \citealt{presotto2014}; \citealt{cattapan2019}) or wavelet-like
  decomposition techniques (e.g., \citealt{ellien2021}).

Indeed, it is not easy to understand whether the observed diffuse
light is bound to the BCG or to the cluster (true ICL), if one uses
only imaging (e.g., \citealt{kluge2021}; \citealt{montes2021}). As
numerical simulations suggest (e.g., \citealt{dolag2010}), the
population of ICL stars should have a higher velocity dispersion than
BCG stars and be more similar to that of the host cluster. This can be
observed in a few cases where detailed spectroscopic studies can be
performed, such as when studying the kinematics of planetary nebulae
or globular clusters around nearby BCGs (e.g.,
  \citealt{arnaboldi1996}; \citealt{longobardi2018};
  \citealt{alamomartinez2021}). However, these tracers can only be
  used for very low redshift objects ($z << 0.1$).  To our knowledge,
there is only one cluster in which ICL occurs at the cluster center
without the presence of a BCG, namely Abell~545
(\citealt{struble1988}; \citealt{barrena2011} and references
therein). This case is clear evidence that diffuse light is bound to
the gravitational potential of the cluster.

Among the mechanisms for ICL formation in galaxy clusters, the
  most important are mass loss during galaxy mergers, tidal disruption
  of dwarf galaxies, tidal stripping of intermediate and massive
  galaxies, and accretion of ICL from groups also called
  preprocessing (see e.g., \citealt{contini2021}).  Numerical
  simulations suggest that half of the ICL comes from galaxies
  associated with the family merger tree of BCGs in clusters (e.g.,
  \citealt{murante2007}) and that stellar stripping by an established
  cluster potential is stronger in the innermost region of the haloes
  and in clusters than in groups (\citealt{contini2018}).  On the
  other hand the mechanisms related to galaxy mergers and
  galaxy-galaxy encounters are favored in the group environment where
  the encounter velocity is low and comparable to the internal
  velocity of stars in galaxies. In fact, numerical simulations show
  as  slow interactions between galaxies in the group environment
lead to strong tidal stripping and the formation of tidal tails and
streams that decay with time and evolve into a more diffuse and
amorphous ICL envelope in the cluster environment
(\citealt{rudick2006}; \citealt{rudick2009}).

According to \citet{rudick2009}, $40\%$ of the cluster ICL is
generated in streams and \citet{contini2014} calculated that
preprocessing can contribute up to $30\%$ of the total ICL in massive
clusters.  The complex substructure of diffuse light in the Virgo
cluster also suggests that ICL is related to the hierarchical nature
of cluster assembly and is not the product of uniform accretion around
a central galaxy (\citealt{mihos2005}).  The analysis of {\em James
  Webb} Space Telescope (JWST) data allows the detection of a lot of
substructures in the ICL of distant clusters suggesting that one can
observe the formation of the diffuse extended component in clusters
(\citealt{montes2022b}).  Despite the presumed importance of the
preprocessing mechanism for ICL formation in clusters, there is little
evidence for ICL in groups. \citet{mihos2016} describes several cases
of loose groups that show little evidence for ICL despite refined
studies. ICL detection has been reported in a few cases (e.g.,
\citealt{castrorodriguez2003}; \citealt{spavone2018};
\citealt{cattapan2019}, \citealt{raj2020}).  On the other hand, ICL
detection is quite common in compact groups such as Stephan's quintet
(\citealt{mendes2001}), Seyfert's sextet (\citealt{durbala2008}), and
several other cases (e.g., \citealt{darocha2005};
\citealt{cortese2006}; \citealt{darocha2008}; \citealt{poliakov2021};
\citealt{ragusa2021}), though not in all compact groups (e.g.,
\citealt{aguerri2006}).

While inspecting an image taken with the Subaru Prime Focus Camera
(Suprime-Cam) and centered on the galaxy cluster
\object{MACS\,J0329-0211} at $z \sim 0.45$ (hereafter MACS0329), we
detected two bright galaxies in the southeastern region, likely in the
foreground of MACS0329 and surrounded by strong diffuse light.  In
addition to the large amount of spectroscopic data in the region of
MACS0329 obtained in the CLASH-VLT project, we performed new optical
observations to obtain new spectroscopic data with the Italian
Telescopio Nazionale {\em Galileo} (TNG) in the region around the two
galaxies. Here we report the discovery of a group of galaxies,
hereafter called \object{GrG\,J0330-0218}, associated with this pair
of bright galaxies.

The paper is structured as follows. We describe the new observations
and all optical data in Sect.~2. We present the selection of  group
members and the redshift catalog in Sect.~3. In Sects.~4 and 5 we present our
results on the group structure, the two dominant galaxies, and the ICL.
We give a brief overview of the study of the large-scale environment of
GrG\,J0330-0218 in Sect.~6.  Section~7 is devoted to the
interpretation and discussion of our results. In Sect.~8, we give a
brief summary and derive our conclusions.  In this work we use
 $H_0 = 70~\ksmpc$ in a flat cosmology with
$\Omega_0=0.3$ and $\Omega_{\Lambda}=0.7$. In the assumed cosmology,
$1\arcmin$ corresponds to  $\sim 160~\kpc$ at the group redshift.  
Recall that the velocities we derive for the galaxies are
line-of-sight velocities determined from the redshift, $V=cz$. Unless
otherwise stated, we report errors with a confidence level (c.l.) of
68\%.

\section{Observations and data}
\label{data}

Figure~\ref{figottico} shows the southeastern region of the
Suprime-Cam image centered on MACS0329 where we detected two bright
galaxies surrounded by diffuse light. These two galaxies are listed by
NED\footnote{http://ned.ipac.caltech.edu/} as the galaxy pair
APMUKS(BJ)\,B032735.21-02283 (\citealt{maddox1990}).  Since these two
galaxies will prove to be the two brightest galaxies of a group in the
foreground of MACS0329, we refer to them as BGGs (BGG1 to the north
and BGG2 to the south).

\subsection{New spectroscopic TNG data}
\label{tng}

\begin{figure}
  \centering
  \resizebox{\hsize}{!}{\includegraphics[width=\textwidth]{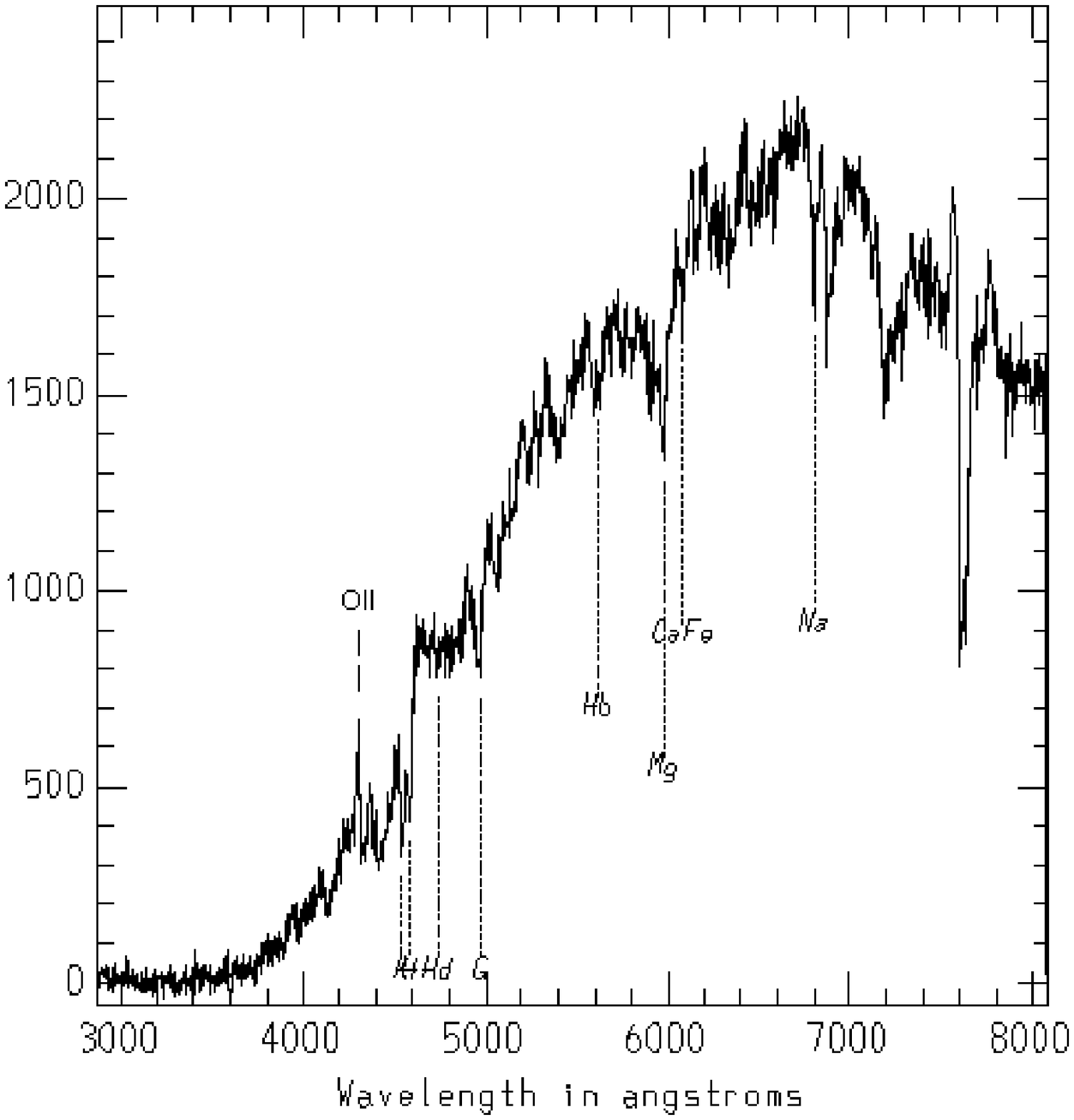}}
  \resizebox{\hsize}{!}{\includegraphics[width=\textwidth]{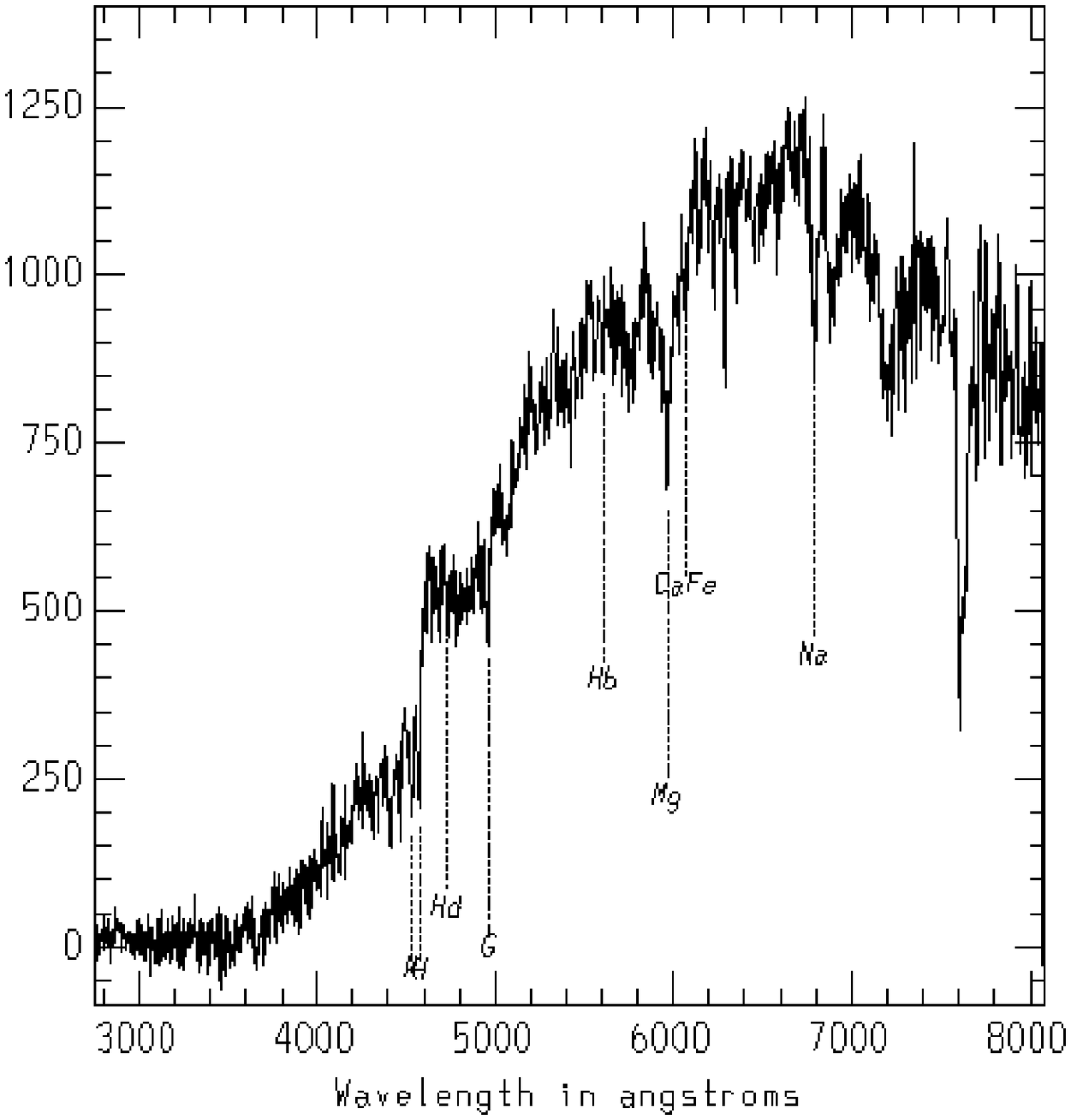}}
    \caption{TNG spectra of BGG1 and BGG2 in the upper and lower panels, respectively. ADU in y-axis.}
    \label{figspectra}
\end{figure}

Long-slit spectroscopic observations of galaxies in the field of
GrG\,J0330-0218 were performed in November and December 2015 and in
January 2016 at the TNG. As instrument, we used the Device
  Optimized for LOw REsolution Spectroscopy (DOLoRes) with the LR-B
grism. With LR-B Grm1 all the relevant spectral features can be
  observed at the group redshift , while the resolution is enough for
  the measurement of accurate galaxy redshifts. In total, we acquired
spectra for 18 galaxies with exposure times between 900 s and 1800
s. These observations focused on the two galaxies BGG1 and BGG2 and on
bright galaxies in their vicinity. The spectra of the two BGGs are
shown in Fig.~\ref{figspectra}.

Spectral reduction and radial velocity estimation were performed using
the standard IRAF\footnote{IRAF is distributed by the National Optical
  Astronomy Observatories, which are operated by the Association of
  Universities for Research in Astronomy, Inc., under a cooperative
  agreement with the National Science Foundation.} tasks and the
cross-correlation technique (Tonry \& Davis 1979). In our experience
with DOLoRes in spectroscopic mode, the nominal velocity errors
provided by the cross-correlation technique must be multiplied by a
factor of 2{--}2.5 (e.g., \citealt{girardi2022}). In this case, comparison
of multiple velocity measurements for four galaxies suggests that we
should be more conservative and allow for errors equal to the nominal
errors multiplied by a factor of 3. In total, we obtained velocity
estimates for 18 galaxies. The median value of the uncertainties in
the velocity measurements is $112~\ks$.

\subsection{CLASH-VLT spectroscopic data}
\label{clash}

The galaxy cluster MACS0329 is part of the ``Cluster Lensing And
Supernova survey with Hubble'' project (CLASH, \citealt{postman2012})
and was surveyed with VIMOS as part of ESO Large Program 186.A-0798
``Dark Matter Mass Distributions of Hubble Treasury Clusters and the
Foundations of LCDM Structure Formation Models'' (PI: P. Rosati).
This program is the panoramic spectroscopic survey of the 13 CLASH
clusters visible from ESO-Paranal, named CLASH-VLT
(\citealt{rosati2014}; Rosati et al.  in prep.). VIMOS was used with
both the low-resolution blue grism (LRb) and the medium-resolution
grism (MR) with typical $cz$ errors of 150 and $75~\ks$, respectively
(\citealt{balestra2016}). Each redshift was also assigned a quality
flag.  Exact details of the CLASH-VLT data reduction can be found in
the study by \citet{mercurio2021}.  The wide-field VLT-VIMOS
spectroscopy was complemented by VLT-MUSE spectroscopy with integral
field in the central cluster regions (\citealt{caminha2019}) with
typical $cz$ errors of $40~\ks$ (\citealt{inami2017}).

For the MACS0329 field, we consider the 1712 galaxies of the CLASH-VLT
catalog with redshift between 0 and 1. Of these, 1637 galaxies are
from VIMOS, 74 from MUSE, and one additional bright galaxy is from a
TNG long-slit observation.  In particular, we only considered VIMOS
redshifts with a quality flag QF$\ge 2$, that is, redshifts with a
reliability $\ga 80\%$.

\subsection{Photometric data}
\label{ima}

We used photometric information from Suprime-Cam data for the MACS0329
field.  Data were retrieved from CLASH
page\footnote{https://archive.stsci.edu/prepds/clash/}, available at
the Mikulski Archive for Space Telescopes (MAST).  A full description
of the reduction of the Suprime-Cam images can be found in the data
section of the CLASH website
\footnote{https://archive.stsci.edu/missions/hlsp/clash/macs0329/data/subaru/}.
Briefly, the images were reduced by one of us using the techniques
described in \citet{nonino2009} and \citet{medezinski2013}.  The total
area covered by the images is $34\times 27~\arcminn$.  Zeropoints are
in the AB system.  In particular, we retrieved the image in the
$R_{\rm C}$ band with an exposure of $2400~\se$ and a depth of
$26.48~\ma$.  The corresponding photometric catalog can be found on
the website CLASH catalogs
\footnote{https://archive.stsci.edu/missions/hlsp/clash/macs0329/catalogs/subaru/}.

We also used data from the Panoramic Survey Telescope and Rapid
Response System (Pan-STARRS, \citealt{flewelling2020}).  In
particular, we extracted $r$, $g$, and $i$ band magnitudes from the
PS1-DR2 data
release\footnote{https://catalogs.mast.stsci.edu/panstarrs}.

\section{Selection of group members and catalog}
\label{memb}

Based on the measured spectra, the two BGGs are very close in velocity
space at $z\sim 0.153$.  The redshift distribution of the 1712
galaxies in the CLASH-VLT redshift catalog was analyzed using the 1D
adaptive-kernel method of Pisani (\citeyear{pisani1993}, hereafter
1D-DEDICA). This distribution reveals a peak of galaxies at $z\sim
0.45$ corresponding to MACS0329, as well as several other foreground
and background structures (Girardi et al. in prep.). Among these
structures, the 1D-DEDICA method detects a foreground peak at
$z=0.153$. It consists of 59 galaxies, 32 of which are located in the
southeastern region of MACS0329.  We added these 59 galaxies from
CLASH-VLT to the 18 galaxies from TNG. Our initial combined redshift
catalog thus consists of 77 galaxies.  Most of the TNG galaxies are
brighter than $R_{\rm C}=20~\ma$, while the data from CLASH-VLT
include galaxies down to $R_{\rm C}\sim24~\ma$.

\begin{figure}
\centering
\resizebox{\hsize}{!}{\includegraphics{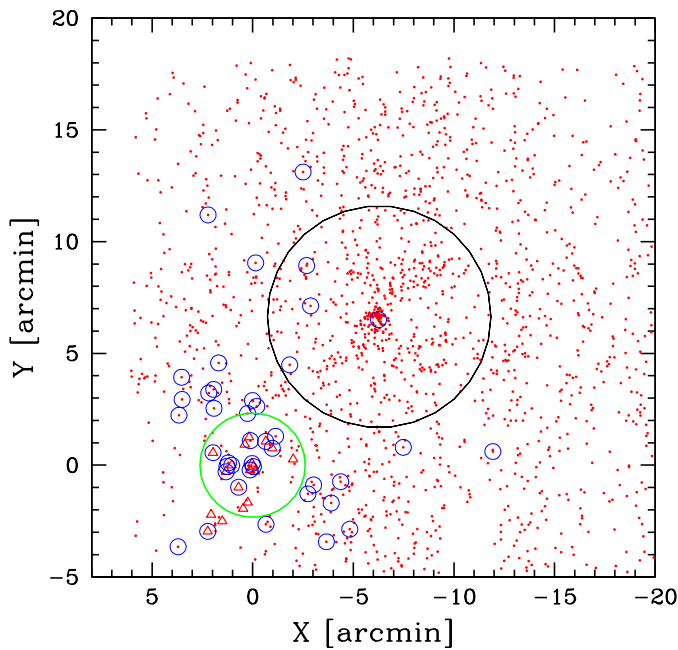}}
\caption
    {Spatial distribution of all galaxies with redshift in the
      MACS0329 field.  The large black circle indicates the
        $R_{200}$ radius of the cluster.  Red triangles and dots show
      galaxies with TNG and CLASH-VLT redshifts, 18 and 1712,
      respectively. The symbols outlined with blue circles indicate
      the 41 group member galaxies. The point 0,0 in the diagram
      indicates the center of GrG\,J0330-0218 and the green
        circle indicates the $0.5~R_{200}$ radius of the group (see
        also Fig.~\ref{figottico}).}
\label{figxycat}
\end{figure}

\begin{figure}
\centering
\resizebox{\hsize}{!}{\includegraphics{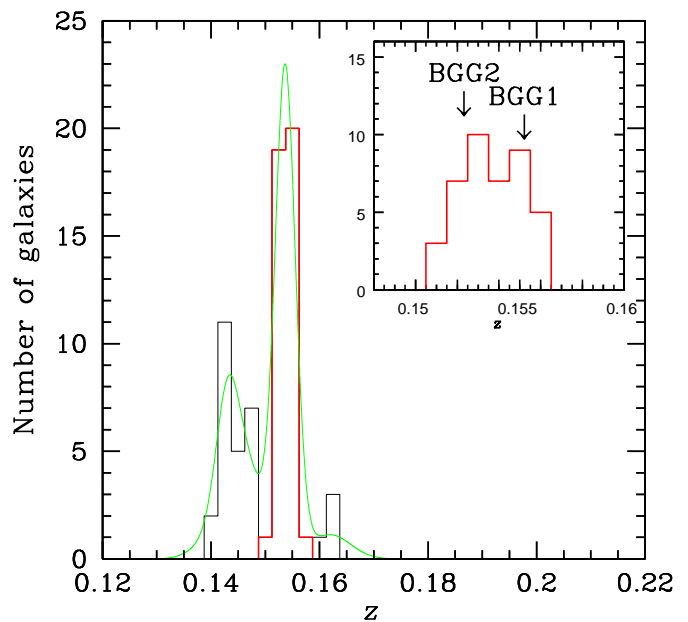}}
\caption
    {Distribution of galaxies in our combined redshift catalog of 77
      galaxies in the range $z=0.12${--}0.22.  The histogram with the
      thick red line refers to the 41 galaxies that belong to the density
      peak of GrG\,J0330-0218, which was identified with the
      1D-DEDICA reconstruction method (faint green line).  The inset
      shows the 41 member galaxies with the redshifts of BGG1 and
      BGG2 indicated.}
\label{fighisto}
\end{figure}

To select group members among the 77 galaxies in our combined redshift
catalog, we used the two-step method known as Peak+Gap (P+G,
\citealt{girardi2015}).  The first step is to apply the 1D-DEDICA
method with adaptive-kernel. With this, we detect a peak at
$z\sim0.1536$ with 41 galaxies in the range $45\,269 \leq{\rm V} \leq
46\,858$ km s$^{-1}$ (see Figs.~\ref{figxycat} and ~\ref{fighisto}). The
nonmembers are 28 foreground and 8 background galaxies. In
particular, there is a foreground peak at $z\sim0.1435$, that is, at more
than $2600~\ks$ in the foreground (cf. Sect.~\ref{LSS}).

In a second step, we combine the space and velocity information in the
``shifting gapper'' procedure (\citealt{fadda1996,girardi1996}). Of
the galaxies that lie within an annulus around the center of the
system, this procedure excludes those that are too far away in
velocity from the main body of galaxies (i.e., are farther away than a
fixed velocity distance called the velocity gap). The position of the
annulus is shifted with increasing distance from the center of the
cluster. The procedure is repeated until the number of cluster members
converges to a stable value.  \citet{fadda1996} suggested a velocity
gap of $1000~\ks$ in the cluster rest-frame and an annulus size of
$0.6~\mpc$ or more to include at least 15 galaxies.  Since the two
BGGs have comparable luminosities, we determine the center of the
group by averaging their positions in right ascension (R.A.) and
declination (Dec.)  [R.A.=$03^{\mathrm{h}}30^{\mathrm{m}}06\dotsec73$,
  Dec.=$-02\degree 18\arcm 24\dotarcs8$ (J2000.0)].

\begin{figure}
\centering
\resizebox{\hsize}{!}{\includegraphics{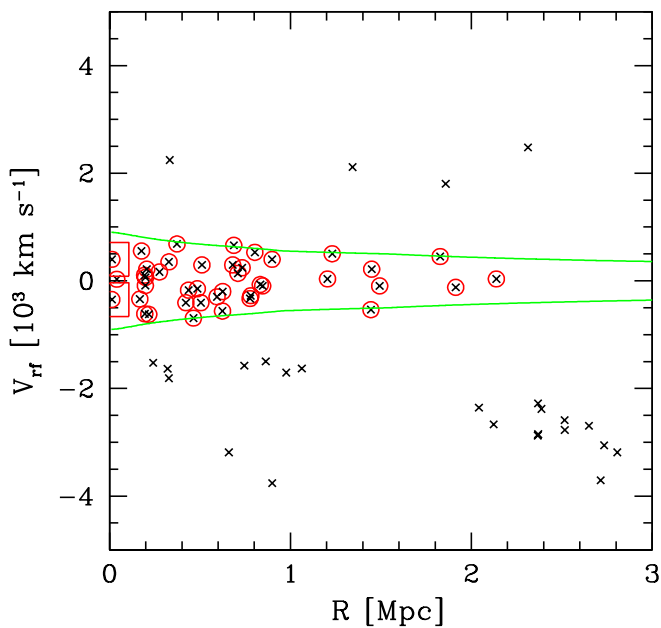}}
\caption
{Rest-frame velocity $V_{\rm rf}=(V-\left<V\right>)/(1+z)$ vs. projected
  group-centric distance $R$ for galaxies with redshifts in the
range   $\pm5000~\ks$ (black crosses). The red circles show
  the 41 members of GrG\,J0330-0218. The large 
  red squares refer to BGG1 and BGG2. The green
  curves contain the region where $|V_{\rm rf}|$ is smaller than the
  escape velocity (see text).  }
\label{figvd}
\end{figure}

Following the procedure described above, the 41 group members are
confirmed. Since a group is expected to have a small velocity
dispersion, we also used the above procedure with a smaller velocity
gap down to $500~\ks$, always confirming the 41 group members.  The
position of the galaxies in the project phase-space is shown in
Fig.~\ref{figvd}.  To highlight the region of the cluster members, we
also plot the escape velocity curves, which were obtained using the
recipe of \citet{denhartog1996}. We used the mass estimate calculated
in the following section and assumed a Navarro-Frenk-White mass
density profile (NFW, \citealt{navarro1997}).  Table~\ref{catalog},
available in full at CDS, lists the velocity catalog for the member
galaxies.

\begin{table}
        \caption[]{Radial velocities of  41 member galaxies of GrG\,J0330-0218}
        \label{catalog}
              $$ 
            \begin{array}{l c c c r c}
            \hline
            \noalign{\smallskip}
            \hline
            \noalign{\smallskip}

\mathrm{ID} & \mathrm{R.A.,Dec.\,(J2000)} & r &V & \Delta V& \mathrm{Source}\\
 & \mathrm{[h:m:s,\degree:\arcmm:\arcs]}&\mathrm{[mag]}&\multicolumn{2}{c}{\mathrm{[km\,s^{-1}]}}& \\
            \hline
            \noalign{\smallskip}  
01& 03\ 30\ 06.85-02\ 18\ 20.3&16.68  & 46528 &153& \mathrm{T}\\
02& 03\ 30\ 06.60-02\ 18\ 28.9&17.01  & 45666 & 93& \mathrm{T}\\
03& 03\ 30\ 07.76-02\ 16\ 06.0&17.15  & 46858 & 75& \mathrm{V}\\
04& 03\ 30\ 21.57-02\ 22\ 03.9&17.36  & 45988 &150& \mathrm{V}\\
05& 03\ 30\ 12.02-02\ 18\ 42.7&18.20  & 45349 & 84& \mathrm{T}\\ 
  \noalign{\smallskip}                                  
            \hline                                          
            \end{array}                                  
            $$
 
            \tablefoot{Full table is available at CDS. We list:
              identification number of each galaxy, ID; right
              ascension and declination, R.A. and Dec.
              (J2000); $r$-band magnitude; heliocentric radial velocity,
              $V=cz$, with error, $\Delta V$; source of spectroscopic
              data (T: TNG, V: VIMOS-VLT, M: MUSE-VLT).  IDs. 01, 02,
            and 03 are  BGG1, BGG2, and BGG3,
            respectively. Magnitudes are generally Pan-STARRS
            Kron-like $r$ magnitudes from PS1 mean data. Those with
            note 'a' are obtained from PS1 stacked data and those with
            'b' are obtained from Suprime $R_{\rm C}$ catalog and
            converted ($r=R_{\rm C}+0.25$, \citealt{fukugita1995}).}
\end{table}


\section{Group structure and properties}
\label{group}

\subsection{Global properties and galaxy population from optical data}
\label{prop}

Analysis of the velocity distribution of the 41 group members was
performed using the biweight estimators for location and scale
included in ROSTAT (statistical routines of \citealt{beers1990}). Our
measurement of the mean redshift of the group is
$\left<z\right>=0.1537\pm0.0001$ (i.e., $\left<V\right>=46\,067\pm41~\ks$).  We estimate the velocity dispersion, $\sigma_{\rm V}$, by
applying the cosmological correction and the standard correction for
velocity errors (\citealt{danese1980}). We obtain $\sigma_{\rm
  V}=369_{-51}^{+20}~\ks$, where the errors are estimated using a
bootstrap technique.

We derive the mass $M_{200}$ within $R_{200}$\footnote{We denote
  $R_{\Delta}$ as the radius of a sphere within which the average mass
  density is $\Delta$ times the critical density at the redshift of
  the galaxy system; $M_{\Delta}$ is the mass contained in
  $R_{\Delta}$.}  using the theoretical relation between $M_{200}$
and the velocity dispersion verified in simulated clusters (Eq.~1 of
\citealt{munari2013}).  To obtain a first estimate of the radius
$R_{200}$ and the group mass $M_{200}$ contained therein, we applied
the relation of \citet{munari2013} to the global value of $\sigma_{\rm
  V}$ which we obtained above.  We considered the galaxies within this
first estimate of $R_{200}$ to calculate a new value for the velocity
dispersion.  The procedure is repeated until we obtain a stable result
that estimates $\sigma_{\rm V,200}=389_{-71}^{+23}~\ks$ for 28
galaxies within $R_{200}=0.77_{-0.14}^{+0.05}~\mpc$.  We derive
$M_{200}=6.0_{-3.9}^{+1.7}~\mtre$.  The uncertainties for $R_{200}$
and $M_{200}$ are calculated using the error propagation of
$\sigma_{\rm V,200}$ ($R_{200}\propto \sigma_{\rm V,200}$ and
$M_{200}\propto \sigma_{\rm V,200}^3$) and an additional uncertainty
of $10\%$ for the mass due to the scatter in the relation of
\citet{munari2013}. The properties of the group are shown in
Table~\ref{tabv}.


\begin{table*}
        \caption[]{Global properties of GrG\,J0330-0218}
         \label{tabv}
            $$
         \begin{array}{c c c c c c c}
            \hline
            \noalign{\smallskip}
            \hline
            \noalign{\smallskip}

            N_{\rm gal} &\mathrm{R.A,Dec.\,J(2000)}&
            \mathrm{<z>}&\sigma_{\rm V}&\sigma_{\rm V,200}&R_{200}&M_{200}\\
            &\mathrm{[h:m:s,\degree:\arcmm:\arcs]}&
            &\mathrm{[km\ s^{-1}]}&\mathrm{[km\ s^{-1}]}&\mathrm{[Mpc]}&[10^{13}M_{\odot}]\\

            \hline
            \noalign{\smallskip}

         41&03\ 30\ 06.73,-02\ 18\ 24.8&0.1537\pm0.0001&369_{-51}^{+20} &389_{-71}^{+23}&0.77_{-0.11}^{+0.05}&6.0_{-3.9}^{+1.7}\\
            \noalign{\smallskip}
            \hline
         \end{array}
         $$

         \end{table*}


Using a number of indicators such as kurtosis, skewness, tail index,
and asymmetry index (\citealt{bird1993}), the analysis of the velocity
distribution shows no evidence of possible deviations from the Gaussian
distribution. We find no evidence of substructure in the 3D
distribution (velocity+positions) using the classical $\Delta$-test of
Dressler \& Shectman (\citeyear{dressler1988sub}) or slightly modified
versions (\citealt{girardi2010}). 

BGG1 has a higher velocity than the mean cluster velocity, as
shown in Fig.~\ref{fighisto} (inset). According to the Indicator test
(\citealt{gebhardt1991}), the velocity of BGG1 is peculiar at the
$>99\%$ c.l.. However, if we consider the two dominant galaxies BGG1
and BGG2 together, that is, their average velocity and not
only the velocity of BGG1, we do not find any
peculiarity.

\begin{figure}
\centering
\resizebox{\hsize}{!}{\includegraphics{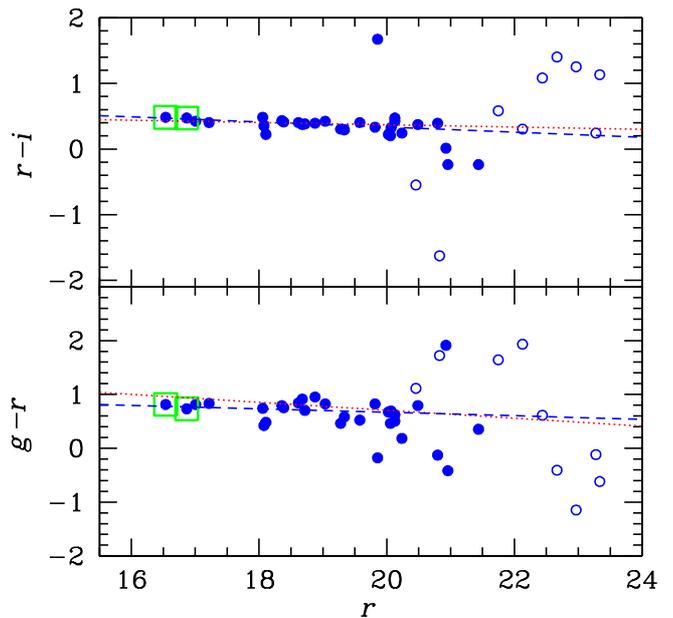}}
\caption
    {Pan-STARRS aperture-color vs. Kron-like magnitude diagrams $r-i$
      vs. $r$ and $g-r$ vs. $r$ in lower and upper
        panels, respectively).  All member galaxies are shown with the
      exception of two very faint galaxies for which PS1 magnitudes are
      not available.  The full circles indicate data  obtained from
      PS1 mean data and open circles those available only as stacked data.
      The large green squares indicate BGG1 and BGG2.  The blue
      dashed lines show the color-magnitude relations obtained for the
      member galaxies. The red dotted lines show the relations
      taken from the literature (see Sect.~\ref{discu}).}
\label{figcm}
\end{figure}

We have also analyzed the position of the group galaxies in the
color-magnitude diagrams.  We can assign Pann-STARRS magnitudes to all
member galaxies, except for two very faint galaxies with $R_{\rm
  C}>23~\ma$.  Figure~\ref{figcm} shows the distribution of member
galaxies in the color-magnitude diagrams $r-i$ vs. $r$ and $g-r$
vs. $r$.  In Fig.~~\ref{figcm} the magnitudes are corrected for
Galactic foreground extinction (\citealt{schlafly2011}). The
color-magnitude relations can be seen down to faint magnitudes $r\sim
20~\ma$.  We use the galaxies with $r<$ 20 and the $2\sigma$ rejection
procedure of \citet{boschin2012twosigma} to fit $r-i=1.115-0.039\times
r$ and $g-r=1.308-0.032\times r$.

Then we calculated the total luminosity of the galaxies
$L_{\rm{gals}}$ within $0.5~R_{200}$. There, the redshift completeness
is about $\sim 75\%$ at the magnitude limit of $19.68=r_{\rm{BGG1}} +
3$. Of the 21 Pan-STARRS galaxies brighter than the magnitude limit,
we have the redshift for 16 galaxies: nine and seven galaxies are
members and nonmembers, respectively. Of the five galaxies with no
spectroscopic redshift, the giant spiral galaxy with a diameter 
  $\sim 0\farcm8$ at the eastern $0.5~R_{200}$ limit is likely a
foreground galaxy (labeled with ``Spi'' in Fig.~\ref{figottico}). We
classify it as a nonmember galaxy. This assumption is also confirmed
by our estimate of the photometric redshift $z_{phot} = 0.11 \pm 0.02$
obtained with the algorithm of \citet{Tarrio2020}.  We thus define two
samples of galaxies: one consisting of the nine spectroscopic members
and one consisting of the nine members plus the four galaxies without
redshift.  We corrected magnitudes for extinction, the K-correction,
and the evolution correction (Eq.~2 in \citealt{girardi2014}), and
summed the luminosities of all galaxies to obtain a range of values
for the global galaxy luminosity within the magnitude limit, $L_{\rm
  obs}$. The contribution from the fainter galaxies is estimated using
the Schechter luminosity function according to the procedure in
\citet{girardi2014}. We obtain
$L_{\rm{gals}}=L_{\rm{obs}}+L_{\rm{faint}}=4.0\times 10^{11}~\lsun$
and $4.6\times 10^{11}~\lsun$ for the two samples of 9 and 13
galaxies, respectively.  A $17\%$ of luminosity is due to the
magnitude corrections and a 19{--}$25\%$ is due to the extrapolation to
faint galaxies. We used the NFW profile to calculate the projected
mass within $0.5~R_{200}$ and obtained a mass-to-light ratio $M/L\sim
130~\ml$.

\subsection{Analysis of X-ray data}
\label{xray}

GrG\,J0330-0218 was serendipitously observed by {\em Chandra} with the
Advanced CCD Imaging Spectrometer (ACIS, \citealt{garmire2003}) within
the field of view of four observations of MACS J0329.6-0211
(Observations ID: 7719, 3257, 3582, 6108). The data set was reprocessed
and cleaned following the steps described in Bartalucci et
al. (\citeyear{bartalucci2017}, see their Appendix~A). We briefly
report the procedures here. Data were processed using 
{\em Chandra} Interactive Analysis of Observations (CIAO,
\citealt{fruscione2006}) version 4.13 using the latest calibration
files updated to version 4.9.6. High-energy particle contamination
was reduced by using the Very Faint
mode\footnote{https://cxc.harvard.edu/cal/Acis/Cal\textunderscore
  prods/vfbkgrnd}.  Observational periods affected by flares were
removed following the procedures of \citet{hickox2006} and
\citet{markevitch2006}. The four observations were merged 
after the cleaning procedure to maximize the statistics. The resulting
total cleaned exposure time is 55 ks. The exposure-corrected image in
the 0.5{--}2.5 keV band was obtained by combining the four observations.

\begin{figure}
\centering
\resizebox{\hsize}{!}{\includegraphics{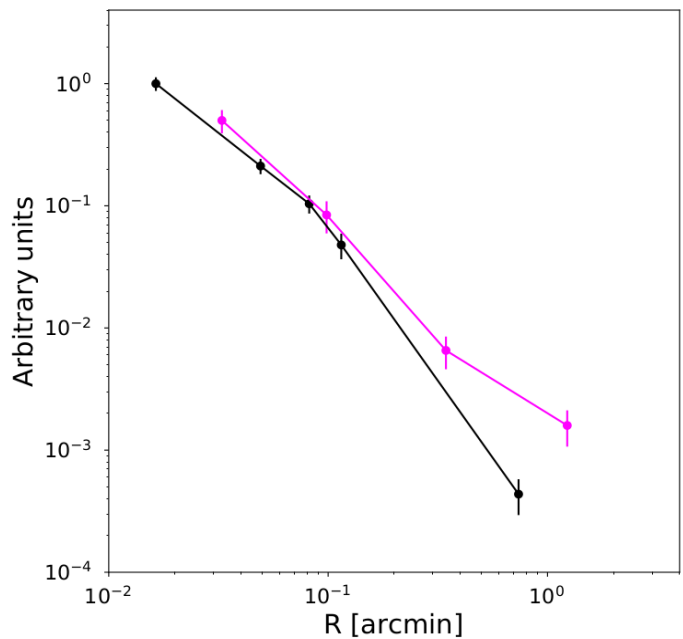}}
\caption
    {Surface brightness profiles of the two X-ray sources. The group
      and point sources are shown by magenta and black lines. 
        Both profiles are extracted in the 0.7{--}2.5 keV band.}
\label{figcomparisonX}
\end{figure}

GrG\,J0330-0218 is in the outermost part of the field of view. The
distance to the aim-point is on average $10\arcmin$, and for this
reason any diffuse emission of the group is expected to be weak. At
the position of GrG\,J0330-0218, the only emission visible on the
image is point-source like and centered on BGG1. We investigated the
nature of this emission by comparing the surface brightness profile
extracted from the group with that of a point source at the same
distance from the aim-point. The profile of the group was extracted
from concentric annuli centered on the BGG1 position, thus considering
both BGG1 and possible intragroup medium. We computed the
background-subtracted and exposure-corrected average surface
brightness profiles in each annulus. The technique of extraction is
detailed in \citet{bartalucci2017}.  The direct comparison of the
background-subtracted and exposure-corrected surface brightness
profiles is shown in Fig.\ref{figcomparisonX}. We scaled the surface
brightness profiles intensity by their maximum value and we applied an
offset of to the radial grids so that the first bins of the profiles
coincide. The group emission is systematically above the point source
at $R>0\farcm2$, $\sim 12\arcsec$, and the point source profile
decreases rapidly.  The $90\%$ average encircled energy fraction at
$1.49~\kev$ is within $\sim 10$ arcsec\footnote{Fig.~4.12 of the {\em
    Chandra} proposal observatory guide
  cxc.harvard.edu/proposer/POG/html/chap4.html}, which is consistent
with the point source profile. The group emission extends to
$1\arcmin$, indicating the presence of a faint extended emission. In
addition, the surface brightness profiles are binned to have at least
$3\sigma$ in each bin.  That is, we have found that the surface
brightness profile extracted from the group is consistent with
extended emission.  The extended emission is centered exactly on the
galaxy BGG1.  We cannot tell whether the observed extended emission is
originated from the diffuse hot gas embedded in the halo of the group
or rather from BGG1 itself.  A dedicate observation is needed to
separate the two components and determine the key thermodynamic
quantities.

\section{Bright group galaxies and diffuse light}
\label{icl}

Inspection of the Suprime-Cam $R_C$ image reveals a clear signature of
diffuse light around the two central galaxies (see
Fig.~\ref{figottico}). In particular, one can see a distinct tail in
the southeastern part of BGG2.  In Sect. \ref{sec:decomposition}, we
first focus on the photometric decomposition of the surface brightness
profiles of the two BGGs to understand the morphological type of these
galaxies. Then, in Sect. \ref{sec:ICL} the ICL is calculated using the
surface brightness cut technique (hereafter SB cut).

\subsection{Photometric decomposition of the two dominant galaxies}
\label{sec:decomposition}

\begin{figure*}
    \centering
    \includegraphics[width=1.\textwidth,angle=00]{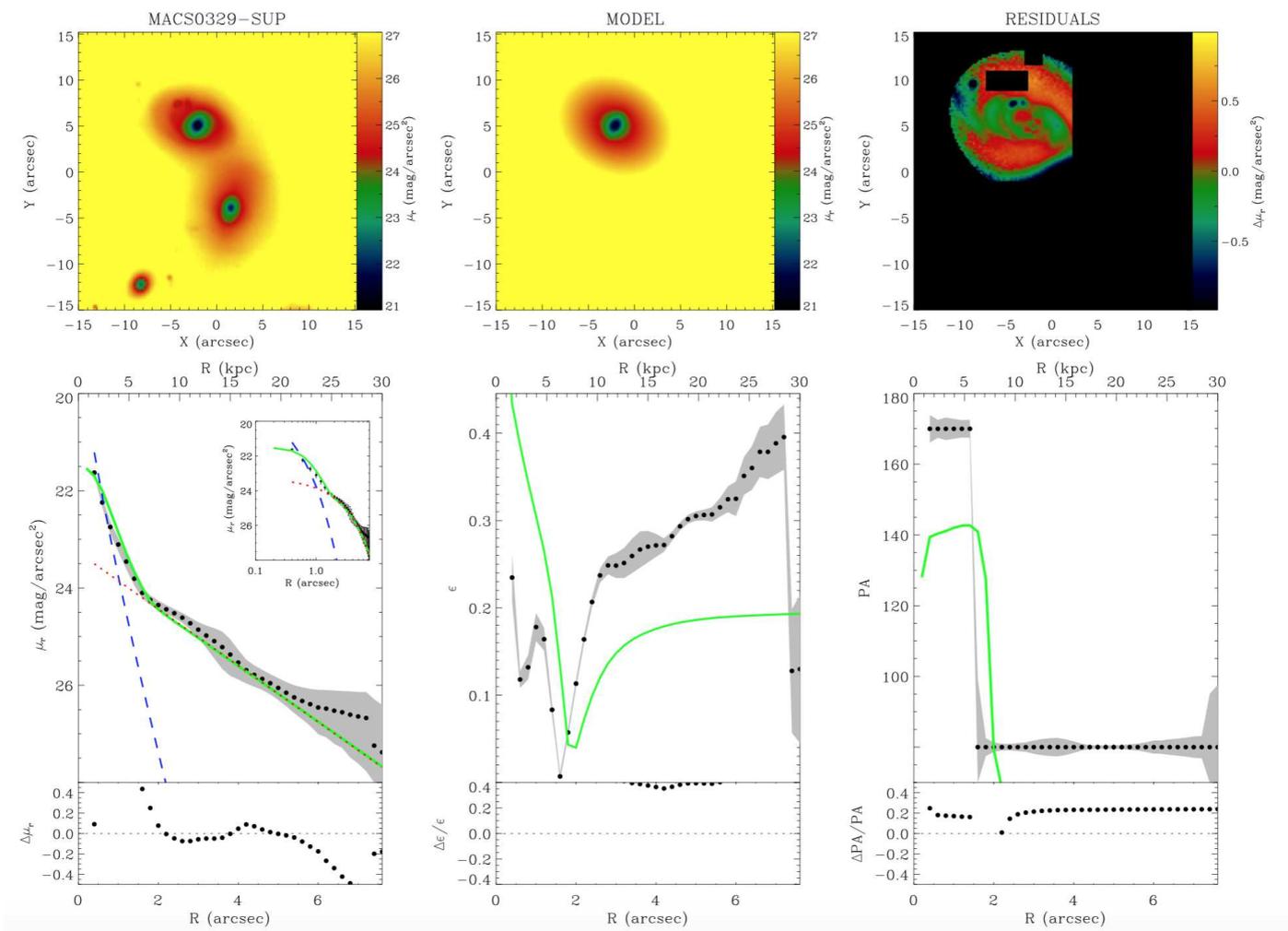}
    \caption{Photometric decomposition of BGG1. The image of the
      region, the model of the galaxy, and the residuals of the
      subtraction of the model from the image are shown in the upper
      panels (left, middle, and right panels).  In the bottom-left
      panel, the ellipse-averaged surface brightness radial profile of
      the galaxy (black dots) and best-fit model (green solid line)
      are shown. Moreover, the light contributions of the bulge 
        (dashed blue line) and disc (dotted red line) is also
      presented. The inset shows a zoom to the central regions of the
      data and the fit using a logarithmic scale for the radial
      distances. In the bottom-central plot the ellipticity profile
      and fit are shown, whereas in the bottom-right panel we present
      the PA profile and fit. The color code of these two last
      plots is the same as for the bottom-left one.}
    \label{fig:decomposition_sup}
\end{figure*}

\begin{figure*}
    \centering
    \includegraphics[width=1.0\textwidth,angle=00]{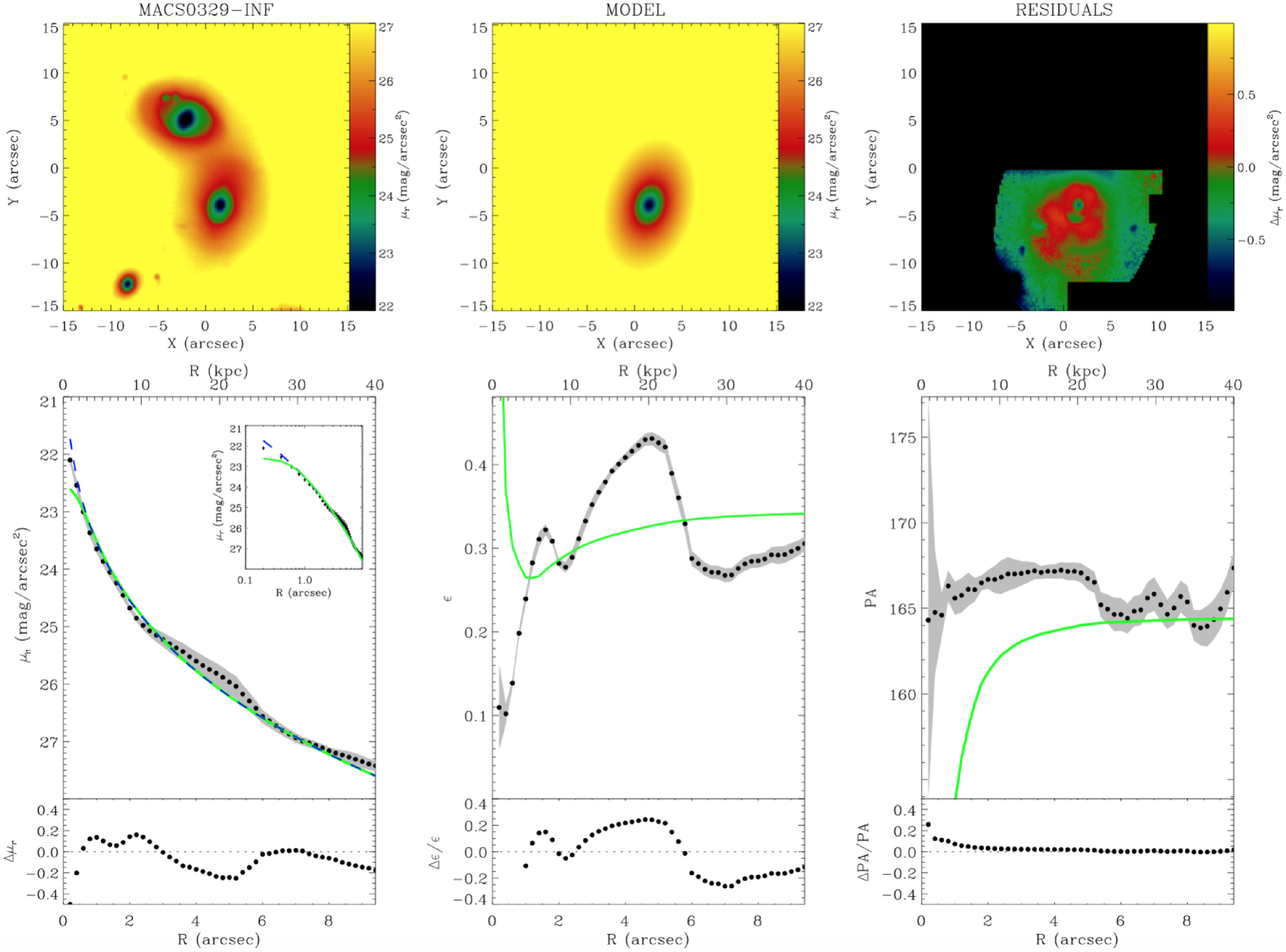}
    \caption{
      Photometric decomposition of BGG2. The panels are the same
      as in Fig.~\ref{fig:decomposition_sup}.
    }
    \label{fig:decomposition_inf}
\end{figure*}

The photometric decompositions were performed using the GASP2D code
(\citealt{mendezabreu2008,mendezabreu2014,mendezabreu2017}). The
method is able to fit several components to the surface brightness
profile of the galaxy. In particular, the latest version of GASP2D is
able to account for a bulge, a disk with no or negative or positive
bending, up to two bars, and a nuclear source to mimic the AGN in the
central region. Details of the method and its configuration can also
be found in \citet{delorenzocaceras2020}.   The PSF is taken into
  account in the decomposition. In particular, we measure the FWHM of
  a set of nonsaturated stars in the image (identified using SDSS
  data) with a Moffat function. The resulting mean PSF is used in
  GASP2D as a kernel to be convoluted with the model of each galaxy,
  thus the final model is seeing-corrected
  \citep{mendezabreu2017}. The sky level was evaluated using the IRAF
  task {\tt imexam}: we measured 10 different (empty) regions of the
  sky, getting for each region the mean value over an area of 25
  pixels. We then computed the average value of the sky from these 10
  measurements. In Figs.~\ref{fig:decomposition_sup} and
\ref{fig:decomposition_inf} we show the results of the photometric
decomposition of BGG1 and BGG2, respectively.

The most interesting result of the decomposition is that BGG1 clearly
has a disk component. Disks are fragile structures, but they are not
uncommon in groups where low relative velocity encounters
occur. Confirmation of the presence of a disk is found in the spectrum
of this BGG1, which is shown in Fig.~\ref{figspectra}. The [OII]
emission line is clearly visible, suggesting that some star formation
is underway. In addition, some spiral arms are visible in the
northeastern side of BGG1, again confirming the presence of a disk in
this galaxy.  We note that a good fit can also be obtained by adding a
bar in the central region, but the data are not sufficient to assess
whether the presence of the bar really improves the fit (e.g., the bar
is real) or if it is an unnecessary component. Therefore, to be
conservative, we opted for the simpler decomposition with only two
components (bulge and disk).  However, it is known that tidal
interactions between two galaxies can trigger the formation of a bar,
especially if both galaxies are massive \citep{noguchi1987}. For this
reason, the possible presence of a bar should not be considered
surprising.  It is also interesting to note that there is a clear tilt
between the position angle (PA) of the bulge and that of the disk. The
bulge is moderately elliptical and perpendicular to the disk component
(PA = 170 degrees for the bulge, PA = 80 degrees for the disk),
another indication that the presence of a bar cannot be completely
ruled out.

On the other hand, BGG2 appears to be a simple early-type
galaxy. The surface brightness profile can be fitted using a single
S\'ersic component with $n=4$.

\subsection{ICL computation}
\label{sec:ICL}

To estimate the amount of ICL in GrG\,J0330-0218 we used the SB
  cut method.  We started with the original image in galaxy counts,
corrected for bias and flat field, and sky subtracted. Then, we ran
Sextractor (\citealt{bertin1996}) over the image to find all objects
in the image (stars and galaxies).  The value of each pixel was
converted to magnitudes using the calibration constant, which was
determined by comparing the unsaturated stars in the Suprime-Cam image
with the same stars from Pan-STARRS (the stars were selected using the
parameter {\tt CLASS\_STAR}).  With the calibration constant in hand,
we can assign a magnitude to each pixel in our image, and knowing that
the pixel size is $0.2$ arcsec, it is easy to determine the area of the
pixel.  With this information, we were able to select all pixels that
were above a certain SB cut.

  Both observations and simulations have shown that this method is
  effective for an ICL surface brightness
  $\mu_V >26.5~\maa$ (\citealt{feldmeir2004}; \citealt{rudick2011};
  \citealt{cui2014}). However, other values are found in the
  literature. For instance, the accurate study of the light profile
  for NGC~1533 in the Dorado group shows that the stellar envelope can
  be detected at about one mag brighter ($\mu_r=25~\maa$,
  \citealt{cattapan2019}).  Using the wavelet
  technique, the mean surface brightness of the ICL component in
  compact groups HCG~79 and HCG~95 are $\mu_R =23.9~\maa$
  and $\mu_R =25.5~\maa$, respectively
  (\citealt{darocha2005}). \citet{furnell2021} suggest a SB cut
  $\mu_B=25~\maa$.

  We ran several tests and decided on a cut at $R_C=25.75~\maa$ ($r
  = 26~\maa$).  We have found that some diffuse (visible to the
  eye) light was cut away, especially in the southeastern stream, when
  fainter values of SB cuts were used.  Adding the surface brightness
  cosmological dimming, $2.5\times (1+z)^4$, and applying the
  K-correction (\citealt{fukugita1995}), the rest-frame SB cut is $R_C
  \sim 25~\maa$ ($r \sim 25.25~\maa$).

\begin{figure}[t]
    \centering
    \resizebox{\hsize}{!}{\includegraphics[width=0.5\textwidth]{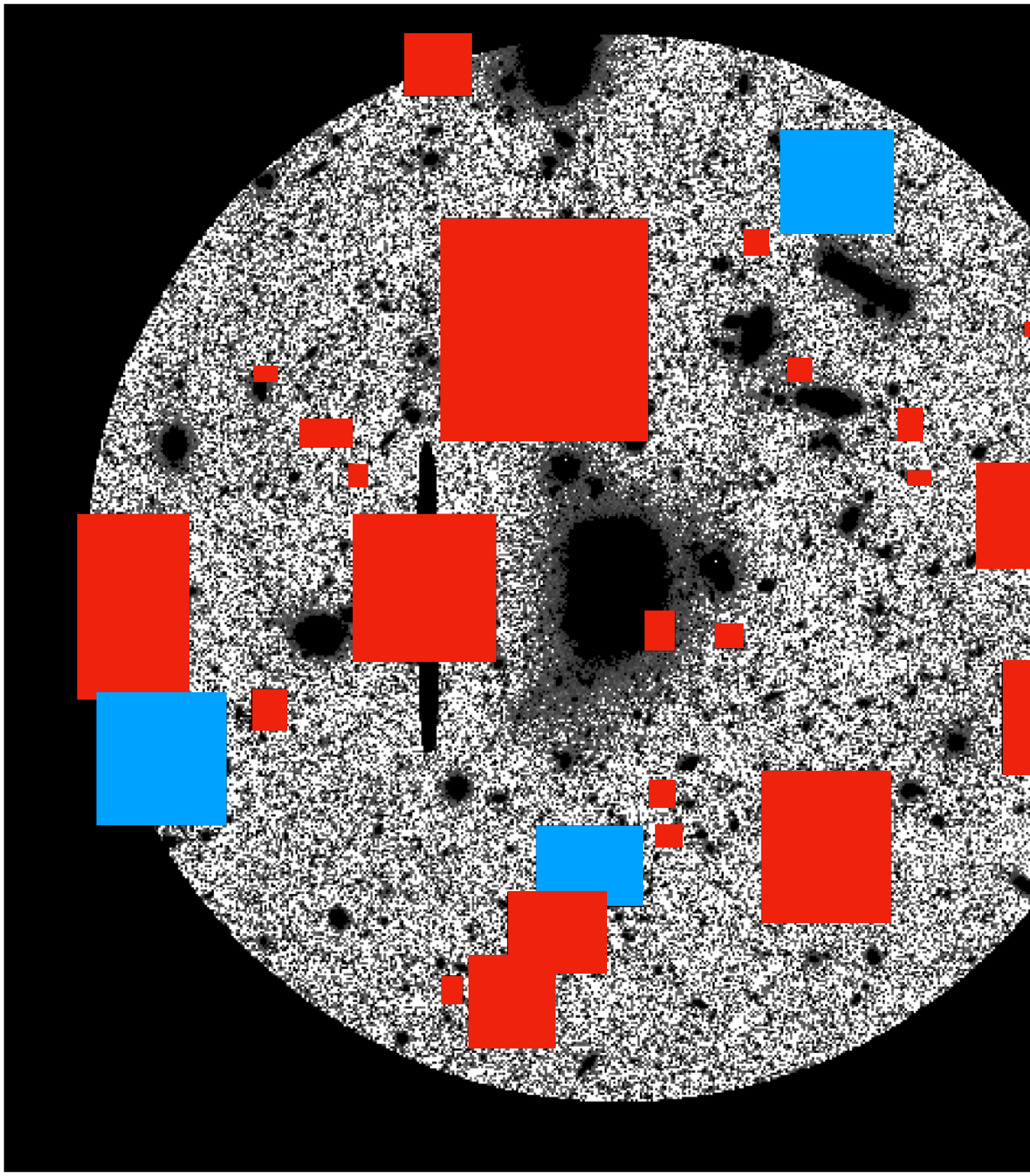}}
 \caption{ Image highlighting the ICL within $0.5~R_{200}$.  Gray
   pixels represent the ICL.  Black pixels are brighter than
   $26~\maa$ and white pixels are fainter than our estimate of the
   sky plus $3\sigma$ (see text). Red regions mask the two bright
   saturated stars and all nonmember galaxies. Light blue regions mask
   bright galaxies without redshift.}
    \label{figSBcut}
\end{figure}

In Fig.~\ref{figSBcut}, we show in gray the pixels that pass through
the cut, that is, pixels fainter than the selected cutoff and brighter
than the mean sky plus 3$\sigma$. All black pixels are brighter than
our SB cut and white pixels are fainter than our sky cut. We calculate
the ICL within $0.5~R_{200}$ and all pixels beyond this radius are
excluded from the computation. We note that we also applied roughly
masks. In particular, for the two very bright saturated stars we
  checked that our masks are large enough to mask all the light from
  the wings of the PSF by measuring the sky just outside the mask
  limits. We computed the mean value of the sky for each of the four
  sides (top, bottom, left, right) using imexam in more than ten points
  per side and we then computed the mean values. We verified that all
  the mean values are below the threshold of the sky plus 3$\sigma$
  that we used as the lower limit for our SB measurements.  We
computed a maximum and a minimum value for the ICL: the former was
computed by applying rectangular masks to the two saturated bright
stars and all the nonmember galaxies (cyan circles and the large
spiral labeled with ``Spi'' in Fig. \ref{figottico}). The
corresponding masks are shown in red in Fig.~\ref{figSBcut}.  The
minimum value was computed by also masking all the galaxies brightest
than $r_{\rm BGG1}+3$ with no available redshift, that is, treating
them as nonmember galaxies (light blue masks in Fig.~\ref{figSBcut}).

The total ICL was then calculated using only the gray pixels of the
image. Specifically, all of these pixels were added together and the
result was converted to an absolute magnitude.  Finally, the total
luminosity of the ICL was calculated.

We find $L_{\rm{ICL}}$ in the range 0.91{--}$0.96 \times
10^{11}\,\lsun$ (in $r$ band).  After applying the same 17\% magnitude
corrections as for $L_{\rm gals}$ (cf. Sect.~\ref{prop}), we can
estimate the fraction of light confined in the ICL with respect to the
total amount of light in the group (galaxies plus ICL) within
$0.5~R_{200}$ as follows:
$f_{\rm{ICL}} = L_{\rm{ICL}}/(L_{\rm{gals}} + L_{\rm{ICL}}) \sim 19${--}22\%.  We
also evaluate the fraction of light in the BGGs plus ICL,
$f_{\rm{ICL+BGGs}} = L_{\rm{ICL+BGGs}}/(L_{\rm{gals}} + L_{\rm{ICL}})
\sim 50${--}56\%.

Our estimate of ICL in GrG~0330-0218 is not expected to be
  affected from the presence of MACS0329. In the cluster rest-frame
  GrG~0330-0218 is projected at $3.2~\mpc$ from the cluster center
  ($\sim$ $1.7~R_{200}$ of MAC0329, \citealt{umetsu2018}) and the whole
  $0.5~R_{200}$ group region here analyzed is out of the cluster $R_{200}$
  (see Fig.~\ref{figxycat}), while ICL is generally reported to be
  important in the cluster core. For instance, \citet{zibetti2005} are
  able to detect ICL out to $700~\kpc$ from the cluster center, but at
  that radius the measured SB is about five $\maa$ fainter
  than at $100~\kpc$.

\section{Large-scale environment}
\label{LSS}

We used NED to analyze the $30\arcmin$ region around GrG\,J0330-0218 and
search for close structures.  About $1\arcmin$ east and $22\arcmin$ southeast
of the group are the two clusters NSC~J033011-021803 and
NSC~J033125-022751, listed by \citet{gal2009} with a photometric
redshift of $z_{\rm phot}=0.1233$. The redshift difference from our
spectroscopic redshift of GrG\,J0330-0218 is comparable to the rms
given by \citet{gal2009}. Thus, it is quite possible that
NSC~J033011-021803 could be the detection of GrG\,J0330-0218 or a
combination of this group and foreground galaxies.  The center of
NSC~J033125-022751 lies outside the region covered by CLASH-VLT and
TNG acquired redshift data.

NED also lists the Zwicky cluster ZwCl~0328.5-0205, $27\arcmin$ northeast
of GrG\,J0330-0218.  In October 2018, using DOLoRes and the LR-B grism
at TNG, we made long-slit spectroscopic observations of the two
dominant galaxies at the center of ZwCl~0328.5-0205
(G1=WISEA~J033116.12-015614.4 and G2=WISEA~J033113.91-015648.2) with
exposure times of $2\times 1800~\se$.  We obtained $cz=43\,277\pm67~\ks$ and
$cz=44\,146\pm48~\ks$, respectively.  The mean redshift of G1 and G2 is
z=0.1458, which can be considered as the redshift of ZwCl~0328.5-0205.
Our analysis of the redshift distribution shows a peak in the
foreground with respect to GrG\,J0330-0218 (see Fig.~\ref{fighisto}).
This peak is located at $z=0.1435$, very close to the redshift of
ZwCl~0328.5-0205.  We conclude that the galaxies forming this peak
probably belong to the outermost regions of this cluster, 
at $\ga 4~\mpc$ from the cluster center.

\section{GrG\,J0330-0218, its nature, and destiny}
\label{discu}

Despite its small mass, GrG\,J0330-0218 appears to be a well-defined
galaxy system.  In the phase-space diagram, the distribution of member
galaxies shows the familiar, characteristic trumpet-shaped pattern
associated with the escape velocity of galaxy clusters, suggesting
that the group is a virialized structure (e.g., \citealt{regos1989};
\citealt{denhartog1996}). The mass-to-light ratio is also consistent
with results on poor clusters (e.g., \citealt{popesso2005}).

The red sequence is visible down to $r\sim 20~\ma$, about 2.5
magnitudes below the value of $M^*$ in the luminosity function (see
Fig.~\ref{figcm}).  Our fitted relations agree well with those
reported in the literature for SDSS bands and for galaxy clusters at
comparable redshift (red dashed lines in Fig.~\ref{figcm}).  The
 $r-i$ vs. $r$ relation is obtained from the results of
\citet{barrena2012}, that is, with the fixed characteristic slope and the
intercept related to the redshift of the system.  The $g-r$
vs. $r$ relation is the one given by \citet{goto2002} in their Fig.~3
for the cluster Abell 1577 at $z\sim 0.14$.  This agreement is
consistent with the fact that galaxies in different environments
appear to show similar red sequences (e.g., \citealt{martinez2010} and
references therein).

As for the X-ray emitting gas, we show the existence of an extended
component centered on BGG1 out to $1\arcmin \sim 160~\kpc \sim
0.2~R_{200}$. In groups, the peak of the X-ray emission usually
coincides with a luminous elliptical or S0 galaxy, which tends to be
the optically most luminous group member. Expected temperature of
  the intragroup medium is $< 2$ keV (e.g., \citealt{mulchaey2000}
and references therein). The fact that the expected temperature
  of the intragroup medium is similar to the temperature of gas in
  galaxies makes difficult to disentangle the intragroup and galaxy
  emissions.  For instance, in GrG\,J0330-0218 we expect a temperature
  in the range 1{--}$2~\kev$ (e.g., Fig.~4 \citealt{lovisari2021} using the
  rescaling $M_{500}=0.75~M_{200}$).  BGG1 has a disk and possibly
  some spiral arms. From the photometric decomposition we obtain the
  bulge-to-total luminosity ratio $B/T\sim 0.3$ which is typical of
  early-type spirals (e.g., Sab{--}Sb according to
  \citealt{oohama2009}). The modest [OII] emission line also excluded
  late-type spirals (cf. our Fig.~\ref{figspectra} {--} upper panel with
  reference spectra by \citealt{kennicutt1992}). For early-type
  spirals one does not expect the X-ray halo typical of a giant
  elliptical, but at most a halo out to few tens of kpc and generally
  at temperature $\la 0.7~\kev$, as obtained from dedicate
  observations (e.g., $20~\kpc$ for Sa Sombrero galaxy by
  \citealt{li2007}; $50~\kpc$ for Sb NGC 6753 by
  \citealt{bogdan2017}). Therefore we would be attempted to explain
  the extended emission in GrG\,J0330-0218 with the presence of the
  intragroup gas, but the scenario is more complicated than this.  In
  fact, the observed X-ray emitting region includes both BGG1 and BGG2
  and these galaxies are in interaction, which is expected to enhance
  the level of X-ray emission (e.g., \citealt{fabbiano2004};
  \citealt{richings2010}).  In groups \citet{desjardins2013} have also
  shows that X-ray emission comes from diffuse features linked to
  individual galaxies likely as a result of the interaction. With the
available data, we cannot separate the intragroup component from the
galaxy emission and need ad hoc data to give any conclusion about
  the existence and properties of the intragroup gas in
  GrG\,J0330-0218.

The central region of GrG\,J0330-0218 is very compact and isolated.
We have verified that it formally meets the photometric criteria for
compact groups (\citealt{hickson1982}). A compact group should have $N
\ge 4$, $\Theta_{N}\ge 3~\Theta_{G}$, and $\mu_{\rm G}<
26.0~\maa$, where $N$ is the number of galaxies within
$3~\ma$ of the brightest galaxy, $\mu_{\rm G}$ is the total magnitude
of these galaxies per arcsec$^2$ averaged over the smallest circle
(with angular diameter $\Theta_{G}$) containing their geometric
centers, and $\Theta_{N}$ is the angular diameter of the largest
concentric circle that contains no other (external) galaxy within the
above magnitude range or brighter. In our case, in the central region
of GrG\,J0330-0218 there are at least four galaxies in the central
region in the required magnitude range (between $m_{\rm BGG1}$ and
$m_{BGG1}+3$). These galaxies are the pair of dominant galaxies and
two close galaxies in the south, the ID.~14 and one nonmember galaxy
(labeled with ``Snm'' in Fig.~\ref{figottico}). This leads to a
surface brightness $\mu_{\rm G} \sim 23.5~\maa$ within a circle of
radius $\Theta_{\rm G}\sim 18\arcsec$.  The bright foreground galaxy in
the north-northeast (labeled with ``Nnm'' in Fig.~\ref{figottico}) is
the closest external galaxy in distance and is located at $\Theta_{\rm
  N}\sim 60\arcsec\, >3~\Theta_{\rm G}$. However, when considering
spectroscopic information, the ``Snm'' galaxy is only projected there,
and thus there are only $N=3$ real three group members.  Of
consequence, the central region of the group can no longer be
classified as a compact group.  Rather, GrG\,J0330-0218 is better
described as a loose group hosting a pair of luminous, dominant
galaxies.

More interesting is the consideration of an ``evolved''
GrG\,J0330-0218, in which the two dominant galaxies have probably
merged into a single, bright central galaxy (hereafter BGG1+2, $r_{\rm
  BGG1+2}=16.08$).  \citet{lin2004} studied the coevolution of
brightest cluster galaxies (BCGs) and host galaxy systems using the
Two Micron All Sky Survey (2MASS).  Considering appropriate
corrections and converting to the 2MASS $K$-band ($r-K=2.5$), we
obtain the $K$-band luminosity of BGG1+2 $L_{K,{\rm BGG1+2}}\sim
4.8\times10^{11}\,\lsun$. This agrees well with the relation proposed
by \citet{lin2004} that a system with $M_{200}\sim 0.6~\mqua$ has a
$L_{K,\rm {BCG}}$ in the range 4.7{--}$5.1 \times 10^{11}~\lsun$ (see
their Eq.~1). Considering the evolved GrG\,J0330-0218, we can also
discuss the possibility that this system can be classified as a fossil
group from a photometric point of view, that is, if the magnitude
difference between the first and the second brightest galaxy within
$0.5~R_{200}$ is $\Delta m_{12}\ge 2~\ma$ (\citealt{jones2003}). The
luminous galaxy BGG3 (ID.~3 with $r=17.15$) to the north, in close
proximity to $0.5~R_{200}$, is crucial for the classification of
GrG\,J0330-0218.  The distance between this galaxy and the center of
the system is $0.373~\mpc$, which should be compared to $0.5~R_{200}$
with $R_{200}=0.77_{-0.11}^{+0.05}~\mpc$.  In practice, the evolved
GrG\,J0330-0218 could be classified as a fossil system with an
uncertainty of $1\sigma$.  This fits well with a scenario in which the
fossil state of galaxy systems is a transitional state, as suggested
by a variety of observational studies (see \citealt{aguerri2021} for a
review).

As for the ICL content, the values reported in the literature are in a
wide interval. The estimate of $f_{\rm{ICL}}$, the fraction of ICL
luminosity to total luminosity, ranges from 0 to 40\% (see Fig.~3 of
\citealt{montes2022}). This wide range is likely due to a real
intrinsic variance in galaxy systems, but also to variations due to
different methods (see our discussion in Sect.~\ref{intro} and, e.g.,
Tab.~1 of \citealt{kluge2021} for a specific comparison).  Other
variations are due to the use of images of different depth and
different definitions of the total luminosity of the system.  All this
makes a direct comparison of our value $f_{\rm{ICL}}\sim 20\%$ with
observational works quite tricky. As an example, we can attempt a
comparison with the results of \citet{kluge2021} who used several
methods and in particular several SB cuts analyzing a sample of 170
clusters. Within the 1-sigma error, our estimate agrees with their
value $f_{\rm{ICL}}=13\pm13\%$ obtained with SB cut at
$26~\gmaa$. Our estimate also fits well among the values computed
in galaxy groups, which, as more massive systems, show a large spread
ranging from 0 to 40\% (see Fig.~9 of \citealt{ragusa2021}).  The
analysis of \citet{martinez2023} of a group at $z\sim 0.2$ shows the
variation in the $f_{\rm{ICL}}$ estimate when using two different
methods. In the $r$ band, the value of $f_{\rm{ICL}}$ goes from $\sim
4\%$ to $\sim 9\%$ when using a brighter cut (our rest-frame cut is
also brighter) to $\sim 30\%$ when using a 2D modeling method. Our
estimate lies within the range.
    
  We can make a more significant comparison using $f_{\rm{ICL+BGGs}}
  =50${--}56\%, which is a more robust quantity than the luminosity of
  ICL and BCG separately. After a small correction of $-5\%$, our
  estimate of $f_{\rm{ICL+BGGs}}$ can be directly compared with the estimates
  obtained by \citet{gonzalez2007} and
  \citet{furnell2021} for 24 and 18 galaxy systems, respectively. The
  $-5\%$ correction is due to the fact that they make computations
  within a larger radius, $R_{500}$ instead of 0.5 $R_{200}$ (see
  Fig.~5 of \citealt{gonzalez2007}). As shown in Fig.~4 of
  \citet{gonzalez2007} and Fig.~13 of \citet{furnell2021},
  $f_{\rm{ICL+BCG}}$ decreases for increasing system mass.
  Considering that GrG\,J0330-0218 has a velocity dispersion of
  $\sigma_{v}\sim 370~\ks$ and mass $M_{200}\sim 6~\mqua$, our value
  of $f_{\rm{ICL+BGGs}}$ agrees well with their results for systems of
  similar mass.  Our estimate also fits in the range of values
  measured for galaxy systems at $z=0.1${--}0.2 (see Fig.~15 of
  \citealt{furnell2021}).

As for the formation of ICL in GrG\,J0330-0218, we note that it can be
compared to one of the prototypical cases described in the numerical
simulations of \citet{rudick2009}, where two massive galaxies interact
outside the cluster environment and produce streams of ICL.  According
to \citet{rudick2009}, the generation of streams is favored by the
strong tidal fields associated with close interactions and mergers
between pairs of galaxies.  These streams are destined to grow very
rapidly and then decay slowly when the interaction occurs outside a
cluster core. However, as accretion-driven growth of large-scale
structure proceeds, the ultimate fate of these streams is to
contribute to diffuse light in clusters. As in GrG\,J0330-0218, when
mapping the surface brightness to deep levels, images of groups are
particularly rich of a lot of features such as streams, tails, or
arclike structures (e.g., \citealt{spavone2018};
\citealt{cattapan2019}).

In the case of GrG\,J0330-0218, we note that it is projected about
$4~\mpc$ from the center of the cluster ZwCl~0328.5-0205 and the velocity
difference is $\Delta V_{rf} < 2100~\ks$.  Since galaxy clusters extend
far beyond $R_{200}$ (e.g., \citealt{biviano2002};
\citealt{rines2013}) and the expected infall velocity between galaxy
clusters can be on the order of a few thousand of $\ks$ (e.g., $3000~\ks$
\citealt{sarazin2002}), GrG\,J0330-0218 may be destined to fall on
the close Zwicky cluster, consistent with theoretical expectations
that ICL produced in groups is destined to form ICL in clusters.

\section{Summary and conclusions}
\label{summa}

We summarize our results on GrG\,J0330-0218, a foreground group
of galaxies that we serendipitously discovered during our visual
inspection of a Suprime-Cam image of the galaxy cluster MACS0329 at $z\sim
0.45$. Our investigation began with the detection of diffuse light
around a pair of bright, comparably luminous galaxies, which were then
identified as the two brightest galaxies in the group, BGG1 to the north
and BGG2 to the south. We used a large redshift sample obtained as part
of the CLASH-VLT project and ad hoc observations from TNG (1712 and
18 galaxies, respectively) to obtain a sample of 41 member
galaxies. Our main results are:

\begin{enumerate}
    \item The mean redshift of the
group is $\left<z\right>=0.1537\pm0.0001$ and the (line of sight)  velocity
dispersion is $\sigma_{\rm V}=369_{-51}^{+20}~\ks$.
    
    \item The distribution of galaxies in the phase space diagram
      indicates that the group is a dynamically evolved system. We
      estimated the dynamical virial mass $M_{200}\sim 6~\mtre$ within
      $R_{200}=0.77~\mpc$.
    
    \item The group is also characterized by a typical value for its
      mass-to-light ratio, and the red sequence is clearly visible and
      extends down to 2.5 magnitudes below the value of $M^*$.
    
    \item We find extended X-ray emission out to $\sim 0.2~R_{200}$
      centered on BGG1.

\item We estimate the fraction of light in the ICL $\sim 20\%$, and
  the fraction included in ICL plus BGGs $\sim 50\%$, which are
  acceptable values within the variance of the values reported in the
  literature.

\end{enumerate}

We would like to emphasize that we were able to discover this group
only thanks to the combination of two factors.  First, we interpreted
the presence of diffuse light around the galaxy pair as a product of
dynamical interaction.  Second, thanks to the huge number of redshifts
available from the CLASH-VLT project, we obtained a first indication
of the presence of a foreground galaxy system in the region around the
galaxy pair.  These two facts have led us to better investigate the
nature of the galaxy pair.

We conclude that it is worthwhile to analyze the bright pairs of galaxies 
surrounded by diffuse light in redshift space to check for the
presence of parent groups. The ICL has the potential to provide a
wealth of information about the formation and evolution of galaxy
systems. Here we propose that galaxy pairs with ICL could be used
as signposts for galaxy groups.
  
 \begin{acknowledgements}
   
    We thank Alfonso Aguerri for useful suggestions on the ICL
    computation and Paula Tarr\'io for the calculation of the
    photometric redshift of the large spiral galaxy located east of
    the center of the group. We thank the referee for his/her useful
    and constructive comments. We acknowledge financial support
    through grants PRIN-MIUR 2017WSCC32, 2020SKSTHZ, INAF-Mainstreams
    1.05.01.86.20. MG acknowledge financial support from the
    University of Trieste through the program FRA 2022.  SZ is
    supported by Padova University grant Fondo Dipartimenti di
    Eccellenza ARPE 1983/2019.
    
    This publication is based on observations made on the island of La
    Palma with the Italian Telescopio Nazionale Galileo (TNG), which
    is operated by the Fundaci\'on Galileo Galilei {--} INAF (Istituto
    Nazionale di Astrofisica) and is located in the Spanish
    Observatorio of the Roque de Los Muchachos of the Instituto de
    Astrof\'isica de Canarias. This publication is based on
    observations collected at the European Southern Observatory under
    ESO Large Programme 186.A-0798. Based in part on data collected at
    Subaru Telescope and obtained from the SMOKA, which is operated by
    the Astronomy Data Center, National Astronomical Observatory of
    Japan. This work makes use of the Pan-STARRS1 Surveys (PS1) and
    the PS1 public science archive which have been made possible
    through contributions by the Institute for Astronomy, the
    University of Hawaii, the Pan-STARRS Project Office, the
    Max-Planck Society and other Institutions.
\end{acknowledgements}

\bibliographystyle{aa}
\bibliography{biblio}

\end{document}